\documentclass[aps,twocolumn,prc,superscriptaddress,showpacs,nofootinbib,floatfix,amssymb,amsfonts,amsmath]{revtex4-1}

\usepackage{graphicx}
\usepackage{epsf}
\usepackage{amsmath,amssymb}
\usepackage{bm}
\usepackage{color}
\usepackage{ulem}

\begin{document}

\title{
$^8$He and $^9$Li cluster structures in light nuclei
}%

\author{Naoyuki Itagaki}
\author{Tokuro Fukui}
\affiliation{
Yukawa Institute for Theoretical Physics, Kyoto University,
Kitashirakawa Oiwake-Cho, Kyoto 606-8502, Japan
}

\author{Junki Tanaka}

\affiliation{
RIKEN Nishina Center for Aaccelerator-Based Science,
2-1 Hirosawa, Wako, Saitama 351-0198, Japan
}

\author{Yuma Kikuchi}
\affiliation{
Tokuyama College of Technology,
Gakuendai, Shunan, Yamaguchi 745-8585, Japan
}

\date{\today}

\begin{abstract}
The possibility of the $^8$He and $^{9}$Li clusters in atomic nuclei is discussed.
Until now 
most of the clusters in the conventional models have been limited to the closures
of the three-dimensional harmonic oscillators, such as $^4$He, $^{16}$O, and $^{40}$Ca.
In the neutron-rich nuclei, however,
the neutron to proton ratio is not unity, 
and it is worthwhile to think about 
more neutron-rich objects with
$N>Z$ as the building blocks of cluster structures.
Here the nuclei with the neutron number six,
which is the subclosure of the $p_{3/2}$ subshell
of the $jj$-coupling shell model, 
are
assumed to be clusters, and thus we study the $^8$He and $^9$Li cluster structures 
in $^{16}$Be ($^8$He+$^8$He),
$^{17}$B ($^8$He+$^9$Li),
$^{18}$C ($^9$Li+$^9$Li),
and
$^{24}$C ($^8$He+$^8$He+$^8$He).
Recent progress of the antisymmetrized quasi cluster model (AQCM) enables us to
utilize $jj$-coupling shell model wave functions as
the clusters rather easily.
It is shown that the
$^8$He+$^9$Li
and
$^9$Li+$^9$Li
cluster configurations cover 
the lowest shell-model states of
$^{17}$B and $^{18}$C, respectively.
To predict the cluster states with large relative distances,
we increase the expectation value of the principal quantum numbers 
by adding the nodes to the lowest states
under the condition that the total angular momentum
is unchanged (equal to $J^\pi =0$).
As a result, developed cluster states are obtained
around the corresponding threshold energies.
The rotational band structure of $^{24}$C, which reflect the symmetry of equilateral triangular 
configuration ($D_{3h}$ symmetry) of three $^8$He clusters, also appears around the threshold energy.

\end{abstract}

\maketitle

\section{Introduction}

The $^4$He nuclei have
been known as $\alpha$ clusters,
which can be subsystems
in some of light nuclei~\cite{Brink,RevModPhys.90.035004}. 
The binding energy of  $^4$He 
is quite large in the light mass region, 
and on the contrary, the relative interaction between $^4$He nuclei is weak.
Therefore
each $^4$He can be considered as a subunit called $\alpha$ cluster.
The candidates for  $\alpha$ cluster structures have been discussed
for many years~\cite{PTPS.68.29},
including the second $0^+$ state of $^{12}$C
with a developed three-$\alpha$ cluster structure
called Hoyle state~\cite{Hoyle}.

In most of the conventional cluster models, the clusters have been limited to the closure
of the three-dimensional harmonic oscillator, such as $^4$He, $^{16}$O, and $^{40}$Ca,
where the contribution of the non-central interactions (spin-orbit and tensor interactions) vanishes.
However, in the nuclear systems, 
the symmetry of the $jj$-coupling shell model
is
more important,
where the contribution of the spin-orbit interaction
breaks the symmetry of the three-dimensional hoarmonic oscillator,
and  the subclosure of
$j$-upper shells,  
$f_{7/2}$, $g_{9/2}$, and $h_{11/2}$,
is essential in explaining
the observed magic numbers of 28, 50, and 126~\cite{Mayer}.
Indeed this spin-orbit interaction is 
also
known as a driving force in breaking the
$\alpha$ clusters~\cite{PhysRevC.70.054307}.
Therefore, 
it would be meaningful to extend the traditional definition of the clusters;
different objects could be candidates for the clusters.

Now we focus on the neutron-rich nuclei, which have been 
the main subject of nuclear structure physics for decades.
In neutron-rich nuclei, the ratio of proton number and neutron number
deviates from unity.
Therefore, it is worthwhile to consider neutron-rich clusters 
whose neutron numbers correspond to the subclosure of the $j$-upper orbits
of the $jj$-coupling shell model,
where the spin-orbit interaction works attractively. 
Here we discuss the possibility that nuclei with the neutron number six,
which is the subclosure of the $j$-upper shell, $p_{3/2}$, can be clusters.
Previously we have discussed the possibility of $^{14}$C cluster
as building blocks of medium-heavy nuclei~\cite{PhysRevC.101.034304},
whose proton number (six) corresponds to the subclosure of $p_{3/2}$.
As the next step, we show the possibility of  the $^8$He (two protons and six neutrons) 
and $^9$Li (three protons and six neutrons) cluster structures
in $^{16}$Be ($^8$He+$^8$He),
$^{17}$B ($^8$He+$^9$Li),
$^{18}$C ($^9$Li+$^9$Li),
and
$^{24}$C ($^8$He+$^8$He+$^8$He).

It has been studied that Be isotopes are well described 
as two $\alpha$ clusters with valence neutrons.
Here, the molecular-orbit structure of the valence neutrons,
where each valence neutron rotates not around only one $\alpha$ cluster but around two $\alpha$ clusters,
has been found to be important~\cite{PhysRevC.61.044306,PhysRevC.62.034301,PhysRevC.65.044302,
PhysRevC.77.067301,PhysRevC.78.011602,PhysRevC.85.014302,10.1093/ptep/ptz169}.
Thus $^4$He+$^6$He and $^5$He+$^5$He configurations
mix for instance in $^{10}$Be. However, it is also known that some of the excited states of $^{12}$Be
has not the molecular-orbit but 
the atomic-orbit structure of the $^6$He+$^6$He or
$^4$He+$^8$He configuration~\cite{PhysRevLett.100.182502,PhysRevC.85.014302,PhysRevC.85.044308},
where each valence neutron sticks to one of the $\alpha$ clusters.
 It should be stressed that the appearance of both molecular and atomic-orbit structures
 in Be isotopes including $^{16}$Be has been systematically studied with
 generalized two-center cluster model (GTCM)~\cite{GTCM}.
In the present study, we treat $^8$He as a cluster based on the atomic-orbit picture and go beyond
the Be isotopes. 

The nucleus $^8$He is the dripline nucleus of the He isotopes
and has a neutron-halo structure.
The valence neutrons are known to have an intermediate character of 
di-neutron structure and independent particle motion~\cite{PhysRevC.74.017307,PhysRevC.78.017306}.
Nevertheless, the two-neutron separation energy of 2.12~MeV is larger 
than that of $^6$He (0.98~MeV), and 
here we simplify it with the $jj$-coupling shell model configuration with the $^4$He core
and consider this nucleus as a subunit.

We also introduce $^9$Li cluster and discuss
$^{17}$B ($^8$He+$^9$Li) and
$^{18}$C ($^9$Li+$^9$Li).
The neutron separation energy of $^9$Li is 4.06~MeV and not very large, 
but this is larger than $^8$He and various $^9$Li+$n$+$n$ models have been applied to
$^{11}$Li so far~\cite{ESBENSEN1992310}.
Although the structure of $^9$Li itself is a subject  to be carefully investigated,
here we simplify it as a cluster using the lowest shell-model configuration
as in the $^8$He case and discuss the cluster structure in the heavy nuclei. 
The two-center-like deformation was predicted in $^{17}$B 
with the 
$^8$He+$^9$Li configuration~\cite{PhysRevC.52.647},
and we further investigate the appearance of more developed cluster states
around the threshold energy.

Recently,
the wave functions of the $jj$-coupling shell model can be easily 
prepared by starting with the cluster model. 
Indeed,  the  antisymmetrized quasi cluster model (AQCM) proposed in 
Refs.~\cite{PhysRevC.94.064324,PhysRevC.73.034310,PhysRevC.75.054309,PhysRevC.79.034308,PhysRevC.83.014302,PhysRevC.87.054334,ptep093D01,ptep063D01,ptepptx161,PhysRevC.97.014307,PhysRevC.98.044306,PhysRevC.101.034304}
allows smooth transformation of the cluster model 
wave functions to the $jj$-coupling shell model ones,
as well as the incorporation of the effects of the 
spin-orbit interaction, which is absent in many of the traditional $\alpha$ cluster models.
Therefore, now we can utilize $jj$-coupling shell model wave functions
as the building blocks of the cluster structure.
In this article, we introduce $^8$He and $^9$Li clusters using AQCM.

In this paper, we discuss that the
$^8$He+$^9$Li
and
$^9$Li+$^9$Li
cluster configurations cover 
the lowest shell-model states of
$^{17}$B and $^{18}$C, respectively.
Also we show the appearance of developed cluster states 
around the corresponding threshold energies
by orthogonalizing to the lowest states.
In addition,
the rotational band structure of $^{24}$C, which reflect the symmetry of equilateral triangular 
configuration ($D_{3h}$ symmetry) of three $^8$He clusters, will be presented.
Although these cluster states are above the neutron-threshold energies, 
they appear around the cluster-threshold energies.

This paper is organized as follows. 
 The framework of AQCM is described in  Sec.~\ref{Frame}.
The results are shown in Sec.~\ref{Results}, where
 $^8$He+$^8$He structure in $^{16}$Be,
 $^9$Li+$^8$He structure in $^{17}$B,
 $^9$Li+$^9$Li structure in $^{18}$C,
 and
 three $^8$He structure in $^{24}$C
 are discussed in {\bf A.}, {\bf B.}, {\bf C.}, and  {\bf D.}, respectively. 
 The conclusions are presented in Sec.~\ref{Concl}.

\section{framework}
\label{Frame}

\subsection{wave function}

We analyze  the $^{8}$He and $^9$Li cluster structures
within the framework of AQCM.
The neutrons of these clusters correspond to the subclosure of $p_{3/2}$ in
the $jj$-coupling shell model, which can be easily 
prepared starting from the  cluster model;  AQCM 
allows the smooth transformation of the cluster model 
wave functions to the $jj$-coupling shell model ones.


In AQCM, each single particle is described by a Gaussian form
as in many other cluster models including the Brink model~\cite{Brink},
\begin{equation}	
	\phi = \left(  \frac{2\nu}{ \pi } \right)^{\frac{3}{4}} 
		\exp \left[-  \nu \left(\bm{r} - \bm{\zeta} \right)^{2} \right] \chi, 
\label{spwf} 
\end{equation}
where the Gaussian center parameter $\bm{\zeta}$
is related to the expectation 
value of the position of the nucleon,
and $\chi$ is the spin-isospin part of the wave function.
For the size paremeter $\nu$, 
here we use $\nu = 0.23$~fm$^{-2}$.

The Slater determinant $\Phi_{SD}$  is constructed from 
these single particle wave functions by antisymmetrizing them,
which is projected to the eigen states of the angular momenta
by numerical integration,
\begin{equation}
\Phi^J_{MK} = {2J+1 \over 8\pi^2} \int d\Omega {D_{MK}^J}^*R(\Omega) \Phi_{SD}.
\end{equation}
Here ${D_{MK}^J}$ is Wigner D-function 
and $R(\Omega)$ is the rotation operator
for the spatial and spin parts of the wave function.
This integration over the Euler angle $\Omega$ is numerically performed.

Next we focus on the Gaussian center parameters
$\{ \bm{\zeta} \}$.
As in other cluster models, here four single particle 
wave functions with different spin and isospin
sharing a common 
$\bm{\zeta}$ value correspond to an $\alpha$ cluster.
This cluster wave function is transformed to
$jj$-coupling shell model based on the AQCM.
When the original value of the Gaussian center parameter $\bm{\zeta}$
is $\bm{R}$,
which is 
real and
related to the spatial position of this nucleon, 
it is transformed 
by adding the imaginary part as
\begin{equation}
\bm{\zeta} = \bm{R} + i \Lambda \bm{e}^{\text{spin}} \times \bm{R}, 
\label{AQCM}
\end{equation}
where $\bm{e}^{\text{spin}}$ is a unit vector for the intrinsic-spin orientation of this
nucleon. 
The control parameter $\Lambda$ is associated with the breaking of the cluster,
and with a finite value of $\Lambda$, the two nucleons with opposite spin orientations 
have the $\bm{\zeta}$ values, which are complex conjugate  with each other.
This situation corresponds to the time-reversal motion of two nucleons.

For the description of $^{8}$He,  at first, di-nucleon clusters are prepared;
in each di-nucleon cluster, two nucleons with opposite spin and same isospin are
sharing a common value for the Gaussian center parameters.
For the proton part, one di-proton cluster is placed at the origin,
which corresponds to the lowest $(0s)^2$ configuration of the shell-model.
For the neutron part,
three di-neutron clusters with equilateral triangular configuration and small
distance ($R$) between them are introduced,
and 
the imaginary parts of the Gaussian center parameters are given
by setting $\Lambda = 1$ in Eq.~(\ref{AQCM}),
which correspond to the subclosure of the $p_{3/2}$ shell~\cite{PhysRevC.94.064324,PhysRevC.87.054334}.
In the actual calculations, $R$ is set to 0.1~fm.
This $^8$He cluster is the eigen state of $J^\pi = 0^+$, and
projection of the total angular momentum of the total system just gives 
the orbital angular momentum of the relative motion 
in the case of $^{16}$Be ($^8$He-$^8$He).

For $^9$Li, one more proton in the $p_{3/2}$ orbit is added.
The Gaussian center parameter of the proton in the $p_{3/2}$ orbit is introduced
in the following way. 
The proton is placed with
a small $x$ component ($R = 0.1$~fm), and the $y$ component is given following
Eq.~(\ref{AQCM}), where the spin orientation is defined along the $z$ axis
(corresponding to spin-up or spin-down proton).
This $^9$Li cluster is the eigen state of $J^\pi = 3/2^-$.

For the calculations of $^{16}$Be, $^{17}$B, and $^{18}$C,
we translationally shift the Gaussian center parameters of the $^8$He and $^9$Li
clusters and place them on the $z$ axis. For $^{24}$C, three
$^8$He clusters are placed to have an equilateral triangular shape.

\subsection{Hamiltonian}

The Hamiltonian consists of the kinetic energy and 
potential energy terms.
The potential energy has central, spin-orbit, and 
Coulomb parts. For the central part, the Tohsaki interaction
(F1 parameter set)~\cite{PhysRevC.49.1814}  
is adopted, which has finite range three-body nucleon-nucleon interaction terms in 
addition to two-body terms. This interaction is designed to reproduce both saturation 
properties and scattering phase shifts of two $\alpha$ clusters. For the spin-orbit part, 
that of  the G3RS interaction~\cite{PTP.39.91}, which is a realistic 
interaction originally  developed to reproduce the nucleon-nucleon scattering phase 
shifts, is adopted. The combination of these two has been  
investigated in detail in Refs.\ \cite{PhysRevC.97.014307,PhysRevC.98.044306}.

\section{Results}
\label{Results}

\subsection{$^8$He+$^8$He cluster structure in $^{16}$Be}

\begin{figure}
	\centering
	\includegraphics[width=5.5cm]{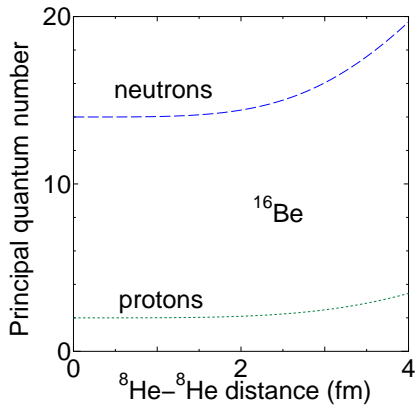} 
	\caption{
Expectation value for the
principal quantum number of the harmonic oscillator ($n$)
for the $0^+$ state of $^{16}$Be
as a function of the distance between two $^8$He clusters. 
Dotted and dashed lines
are results for the
protons and neutrons, respectively.
}
\label{be16-n}
\end{figure}

We start the discussion with the
$^8$He+$^8$He cluster structure in $^{16}$Be.
The dripline nucleus of the Be isotopes is $^{14}$Be,
thus $^{16}$Be
is located outside the neutron dripline.
Experimentaly, the two-neutron separation energy is
$-1.35$~MeV (unbound),
but the ground state is lower than the $^8$He+$^8$He threshold
by 5.77~MeV.

First, we show that the $^8$He+$^8$He model space
covers the lowest shell-model configuration of $^{16}$Be.
The expectation value for the principal quantum number of the harmonic oscillator ($n$)
for the $0^+$ state of $^{16}$Be is shown in Fig.~\ref{be16-n}
as a function of the distance between two $^8$He clusters. 
The dotted and dashed lines
represent the results for the
protons and neutrons, respectively,
which converge to 2 and 14
at small relative distances.
These values agree with the ones for the lowest shell-model configuration;
for the protons, two are in the lowest $0s$-shell and two are in the $p$-shell
($2 \times 1 = 2$), and for the neutrons,
two are in the lowest $0s$-shell, six are in the $p$-shell,
and four are in the $sd$-shell 
($6 \times 1 + 4 \times 2 = 14$).
Therefore, the lowest shell-model configuration is included
in the model space.
With increasing the relative distance between the two $^8$He clusters, 
the components of higher shells mix,
and the $n$ value
rapidly increases.

\begin{figure}
	\centering
	\includegraphics[width=5.5cm]{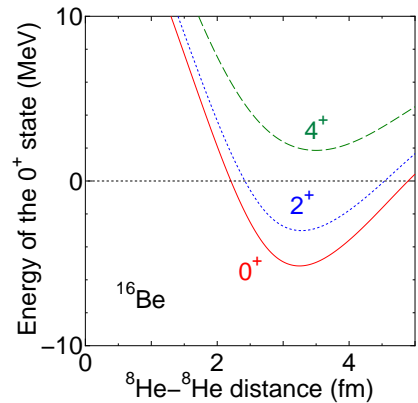} 
	\caption{
Energy curves for $^{16}$Be
measured from the two-$^8$He threshold
as a function of the distance between two $^8$He clusters. 
Solid, dotted, and dashed lines correspond to
the $0^+$, $2^+$, and $4^+$ states, respectively.
}
\label{he8-he8}
\end{figure}

Next, the energy curves for  $^{16}$Be
measured from the two-$^8$He threshold
is shown in Fig.~\ref{he8-he8}
as a function of the distance between two $^8$He clusters. 
The solid, dotted, and dashed lines 
are the results for
the $0^+$, $2^+$, and $4^+$ states, respectively.
It can be seen that the optimal energy 
for the $0^+$ state is obtained
with the relative distance of $\sim3$~fm.
This means that the lowest energy is obtained not at the limit of the shell-model,
and clustering is found to be important,
which is indeed higher shell mixing in terms of the shell model.
Although we do not have adjustable parameters,  the lowest energy  is close
to the experimental binding energy of 5.77~MeV.
Despite this binding energy, which is enough large,
the optimal distance is large owing to the Pauli blocking effect
at short relative distances between two $^8$He clusters,
and developed cluster structure appears.
For the $2^+$ and $4^+$ states,
the optimal distances are slightly larger 
than the one for the $0^+$ state
due to the centrifugal force.

\subsection{$^9$Li+$^8$He cluster structure in $^{17}$B}

\begin{figure}
	\centering
	\includegraphics[width=5.5cm]{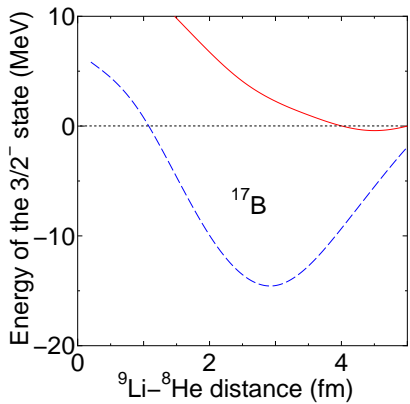} 
	\caption{
Energy for the $3/2^-$ state of $^{17}$B
measured from the $^9$Li-$^8$He threshold
as a function of the distance between $^9$Li and $^8$He
(dashed line). 
Energy curve for the $3/2^-$ state 
orthogonal to 
the state having the optimal
distance of 3~fm (solid line).
}
\label{li9-he8}
\end{figure}

We add one proton and show the result of
$^9$Li-$^8$He cluster configuration in $^{17}$B.
The dashed line in Fig.~\ref{li9-he8}
shows the energy for the lowest $3/2^-$ state of $^{17}$B
measured from the $^9$Li-$^8$He threshold
as a function of the distance between $^9$Li and $^8$He.
Experimentally, the $^{17}$B nucleus is bound from the $^9$Li+$^8$He
threshold by 12.86~MeV. It is not perfect but the obtained 
lowest energy is fairly close to this value.
Similarly to the $^8$He+$^8$He case, 
despite this large binding energy,
the relative distance at the optimal energy is also large owing to the Pauli blocking effect
at short relative distances between the two clusters. 
In this calculation, $^9$Li and $^8$He clusters are placed
on the $z$-axis, while the last proton in $^9$Li stays
on the perpendicular plane, and there is no additional excitation 
to higher shells for this proton when the $^8$He cluster approaches.
Therefore, again the lowest
shell-model configuration of $^{17}$B is included in the model space.

The energy curve for the $3/2^-$ state 
orthogonal to  
the lowest state (relative distance of 3~fm) is shown as the solid line.
It is intriguing to see that the energy minimum point appears
around the threshold energy with a very large relative distance 
of 4~fm.
The appearance of very developed cluster structure around the threshold
is expected, and adding neutrons to this state
 and investigating the molecular-orbital structure
would be performed in the near future.

\subsection{$^9$Li+$^9$Li cluster structure in $^{18}$C}

\begin{figure}
	\centering
	\includegraphics[width=5.5cm]{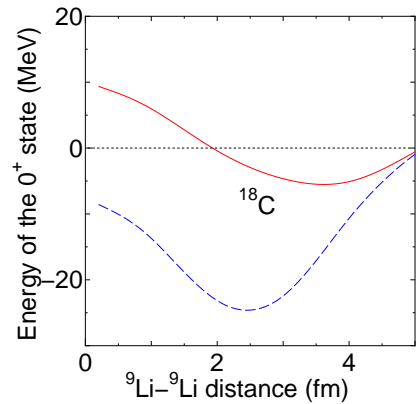} 
	\caption{
Energy for the $0^+$ state of $^{18}$C
measured from the two-$^9$Li threshold
as a function of the distance between two $^9$Li clusters
(dashed line). 
Energy curve for the $0^+$ state 
orthogonal to 
the state having the optimal
distance of 2.5~fm (solid line).
}
\label{li9-li9}
\end{figure}

For $^{18}$C, we introduce a $^9$Li+$^9$Li model.
Here the spin directions of the valence protons in two $^9$Li clusters
are introduced to be anti-parallel, and two valence protons occupy 
the time-reversal orbits.
Thus again the model space
covers the lowest shell-model configuration of $^{18}$C. 
The dashed line in Fig.~\ref{li9-li9} shows
the energy for the lowest $0^+$ state of $^{18}$C
measured from the two-$^9$Li threshold
as a function of the distance between two $^9$Li clusters.
The optimal energy is obtained around the relative distance of 2.5~fm.
Experimentally, the ground state of $^{18}$C is lower than 
the two-$^9$Li threshold by 24.99~MeV.
Again, although we do not use any adjustable parameter,
the obtained optimal energy is fairly close to this value.

The solid line in Fig.~\ref{li9-li9} shows the energy curve for the $0^+$ state 
orthogonal to 
the lowest state with the relative distance of 2.5~fm.
Again, the appearance of the significanly clusterized state around the threshold energy 
with the relative distance of $\sim 4$ fm is expected. 
The intrinsic densities of this state on the $xz$-plane $(y = 0)$ 
with the relative distance of 4~fm are represented by Fig.~5(a) and Fig.~5(b) for protons and neutrons, respectively. As a future work, adding neutrons to this state would be interesting.

\begin{figure}
	\centering
	\includegraphics[width=8cm]{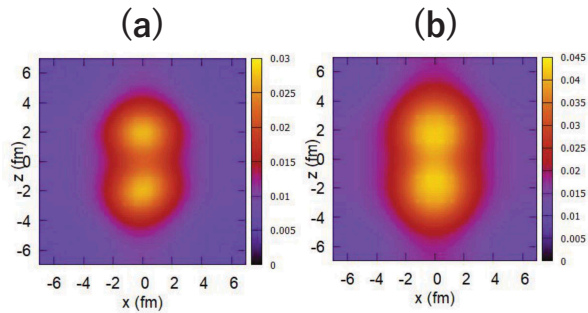}  
	\caption{
Intrinsic density of $^9$Li-$^9$Li 
with the relative distance of 4~fm
on the $xz$-plane (fm$^{-3}$).
(a): protons, (b): neutrons.
}
\label{Int-dens}
\end{figure}

\subsection{three $^8$He cluster structure in $^{24}$C}

\begin{figure}
	\centering
	\includegraphics[width=5.5cm]{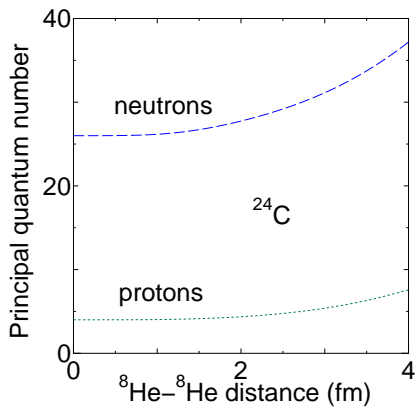} 
	\caption{
Expectation value for
the principal quantum number of the harmonic oscillator 
for the $0^+$ state of $^{24}$C
with the equilateral triangular configuration of three $^8$He clusters
as a function of the distance between two $^8$He clusters. 
Dotted and dashed lines
correspond to the result 
for the protons and neutrons, respectively.
}
\label{c24-n}
\end{figure}

Finally, we discuss the three $^8$He cluster
structure in $^{24}$C.
The dripline nucleus of the C isotopes is $^{22}$C 
and hence $^{24}$C is
beyond
the neutron dripline.
Nevertheless, the three-$^8$He states are shown to appear
around the threshold energy.
First, we show that the three $^8$He configuration with the equilateral triangular 
configuration covers the model space of the lowest shell-model.
The expectation value for the principal quantum number of the harmonic oscillator 
for the $0^+$ state of $^{24}$C
with the equilateral triangular configuration of three $^8$He clusters
is shown in Fig.~\ref{c24-n}
as a function of the distance between two $^8$He clusters. 
Here the dotted and dashed lines 
represent the results for
the protons and neutrons, respectively.
They converge to 4 and 26
at small relative distances, respectively.
These values are ones for the lowest shell-model configuration;
for the protons, two are in the lowest $0s$-shell and four are in the $p$-shell
($4 \times 1 = 4$), and for the neutrons,
two are in the lowest $0s$-shell, six are in the $p$-shell,
and ten and in the $sd$-shell 
($6 \times 1 + 10 \times 2 = 26$).
Therefore, surprisingly enough, 
the lowest shell-model configuration of $^{24}$C
is included within the three-$^8$He model with
an equilateral triangular configuration.
However, experimentally,
the three-$^8$He threshold is located
quite high (more than $E_x = 25$~MeV) 
in the excitation energy,
thus the three-$^8$He configuration corresponds 
to a highly excited state.

\begin{figure}
	\centering
	\includegraphics[width=5.5cm]{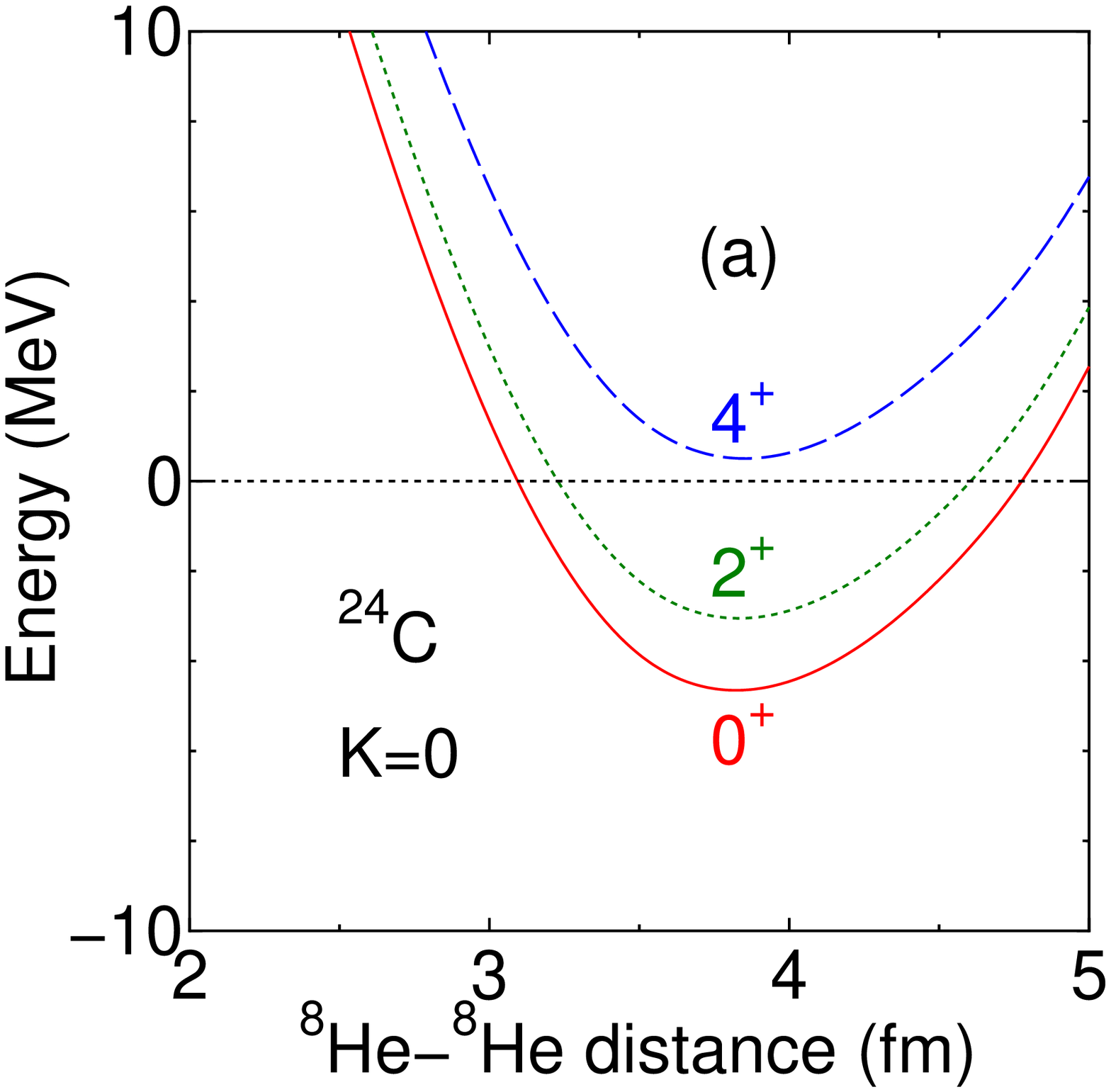} 
	\includegraphics[width=5.5cm]{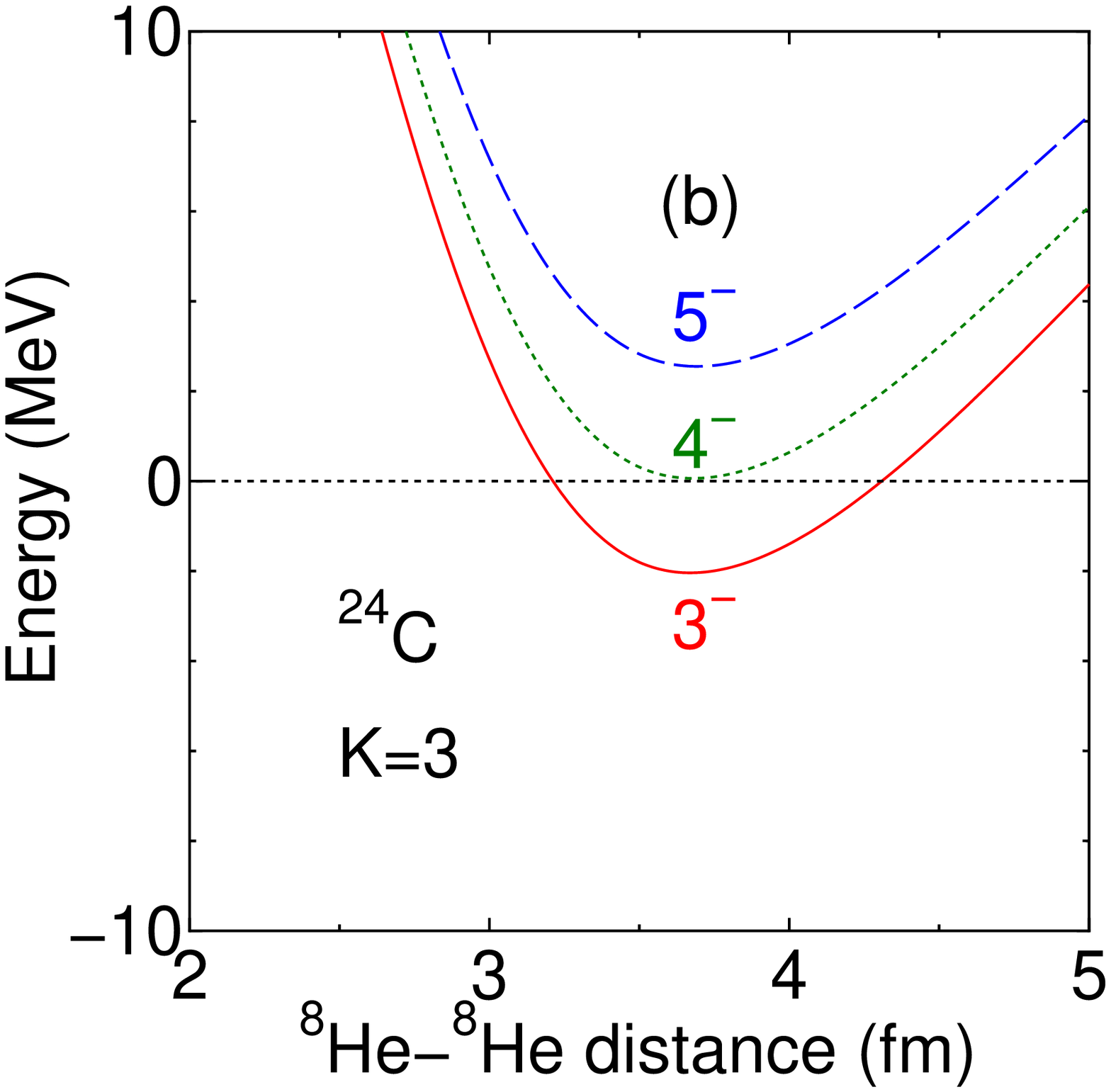} 
	\caption{
Energy curves of $^{24}$C
with the equilateral triangular configuration of three $^8$He clusters
measured from the three-$^8$He threshold
as a function of the distance between two $^8$He clusters. 
(a): positive-parity $K=0$ band, $0^+$, $2^+$, and $4^+$, and
(b): negative-parity $K=3$ band, $3^-$, $4^-$, and $5^-$.
}
\label{c24-eng}
\end{figure}

In Fig,~\ref{c24-eng},
the energy curves of $^{24}$C
with the equilateral triangular configuration of three $^8$He clusters
measured from the three-$^8$He threshold
are shown
as a function of the distance between two $^8$He clusters. 
Here Fig.~\ref{c24-eng}(a) displays the result for the positive-parity
states with $K=0$ ($0^+$, $2^+$, and $4^+$), and 
those for the negative-parity states with $K=3$ ($3^-$, $4^-$, and $5^-$)
can be found in Fig.~\ref{c24-eng}(b).
We can see that both bands appear around the three-$^8$He threshold energy
with large relative distance.

It has been known that if the system has
the equilateral triangular configuration ($D_{3h}$ symmetry),
both $K=0$ ($0^+$, $2^+$, $4^+$ $\cdots$) and
$K=3$ ($3^-$, $4^-$, $5^-$ $\cdots$) rotational bands are possible.
The appearance of these rotational bands has been extensively discussed
in $^{12}$C~\cite{PhysRevLett.113.012502},
which is the signature of the equilateral triangular symmetry of the three $\alpha$ clusters.
Now the $\alpha$ clusters are replaced with the $^8$He
clusters and what we discuss here is considered to be the
neutron-rich version of the $D_{3h}$ symmetry.

\begin{figure}
	\centering
	\includegraphics[width=5.5cm]{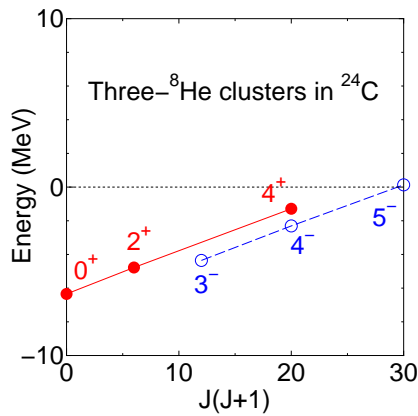}  
	\caption{
Rotational band structure of three-$^8$He
configuration.
Solid and dashed lines
are the results for
the $K=0$ (positive-parity, $0^+$, $2^+$, $4^+$ $\cdots$)
and $K=3$ (negative parity, $3^-$, $4^-$, $5^-$ $\cdots$) bands, respectively.
}
\label{c24-rb}
\end{figure}

The energy eigen states of the three-$^8$He cluster states are obtained by superposing the 
Slater determinants with different relative distances and
diagonalizing the Hamiltonian based on the generator coordinate method (GCM).
The rotational band structure of three-$^8$He
configuration is shown in Fig.~\ref{c24-rb},
where the solid and dashed lines 
correspond to the result for $K=0$ (positive-parity, $0^+$, $2^+$, $4^+$ $\cdots$)
and $K=3$ (negative parity, $3^-$, $4^-$, $5^-$ $\cdots$) bands, respectively.
Two rotational band structures appear around the threshold energy
and they have similar slopes as a function of $J(J+1)$.
   
\section{Conclusions} 
\label{Concl}

Most of the conventional clusters, so far investigated, have been limited
to the closure
of the three-dimensional harmonic oscillator, such as $^4$He, $^{16}$O, and $^{40}$Ca.
Here we discussed the possibility that nuclei with the neutron number six,
which is the subclosure of the $p_{3/2}$ subshell of the $jj$-coupling shell model, 
can be clusters;
the $^8$He and $^9$Li cluster structures
have been  investigated in $^{16}$Be ($^8$He+$^8$He),
$^{17}$B ($^8$He+$^9$Li),
$^{18}$C ($^9$Li+$^9$Li),
and
$^{24}$C ($^8$He+$^8$He+$^8$He).

We have shown that the lowest principal quantum numbers 
of $^{16}$Be, $^{17}$B, $^{18}$C, and $^{24}$C
can be covered within this model.
We have just adopted Tohsaki interaction,
which has finite-range three-body terms,
and there is no adjustable parameter in the Hamiltonian.
Nevertheless the optimal energies
of these nuclei measured from the corresponding threshold energies
are fairly close to the experimental values.
By orthogonalizing the wave functions
to the lowest states, very developed cluster 
states were obtained around the corresponding threshold energies
in $^{17}$B and $^{18}$C.

The appearance of 
$K=0$ ($0^+$, $2^+$, $4^+$ $\cdots$) and
$K=3$ ($3^-$, $4^-$, $5^-$ $\cdots$) 
rotational bands has been extensively discussed
in $^{12}$C~\cite{PhysRevLett.113.012502},
which is the proof for the equilateral triangular symmetry of the three $\alpha$ clusters.
In this study, we have replaced the $\alpha$ clusters with the $^8$He
clusters and  shown the 
neutron-rich version of the rotational band structures 
for the configuration reflecting the
$D_{3h}$ symmetry.
The energy eigen states of the three-$^8$He cluster states are obtained by superposing the 
Slater determinants with different relative distances and
diagonalizing the Hamiltonian.
The two rotational band structures of three-$^8$He
configuration appear around the threshold energy
and they have similar slopes as a function of $J(J+1)$.

As future works, the appearance of the molecular-orbital structure
will be studied 
by adding neutrons to the developed cluster states
obtained here in $^{17}$B and $^{18}$C. 
Also, we investigate the role of the obtained cluster states, 
including resonances above the cluster emission threshold around the Gamow window, in the nuclear reactions including the Big Bang nucleosynthesis.

\begin{acknowledgments}
The authors would like to thank fruiteful discussions with Dr.~Masaki Sasano (RIKEN).
The numerical calculations have been performed using the computer facility of 
Yukawa Institute for Theoretical Physics,
Kyoto University. 
\end{acknowledgments}

\bibliography{biblio_ni.bib}

\end{document}